\documentclass{PoS}

\title{Status of the Cherenkov Telescope Array's Large Size Telescopes}

\ShortTitle{Status of CTA LST}

\author{Juan Cortina\\
  Institut de Fisica d'Altes Energies, Cerdanyola del Valles, E-08193, Spain\\
  E-mail: \email{cortina@ifae.es}}

\author{\speaker{Masahiro Teshima}\\
  Max-Planck-Institut f\"ur Physik, M\"unchen, D-80805, Germany\\
  E-mail: \email{mteshima@mppmu.mpg.de}}

\author{and the LST team for the CTA consortium\thanks{Full consortium list at 
http://cta-observatory.org}}

\abstract{
The Cherenkov Telescope Array (CTA) observatory, will
be deployed over two sites in the two hemispheres. Both sites will be
equipped with four Large Size Telescopes (LSTs), which are crucial to 
achieve the science goals of CTA in the 20-200 GeV energy range. Each LST 
is equipped with a primary tessellated mirror dish of 23 m diameter,
supported by a structure made mainly of carbon fibre reinforced plastic tubes
and aluminum joints. This solution guarantees light weight (around
100 tons), essential for fast repositioning to any position in the sky
in $<$20 seconds. The camera is composed of 1855 PMTs and
embeds the control, readout and trigger electronics.
The detailed design is now complete and production of the first LST,
which will serve as a prototype for the remaining seven, is well underway.
In 2016 the first LST will be installed at the Roque de los Muchachos Observatory on
the Canary island of La Palma (Spain).
In this talk we will outline the technical solutions adopted to fulfill
the design requirements, present results of element prototyping and
describe the installation and operation plans.
}

\FullConference{The 34th International Cosmic Ray Conference,\\
		30 July- 6 August, 2015\\
		The Hague, The Netherlands}

\begin{document}

\section{Overview of the telescope technical design}

\begin{figure}[!htb]
	\centering
		\includegraphics[width=0.8\textwidth]{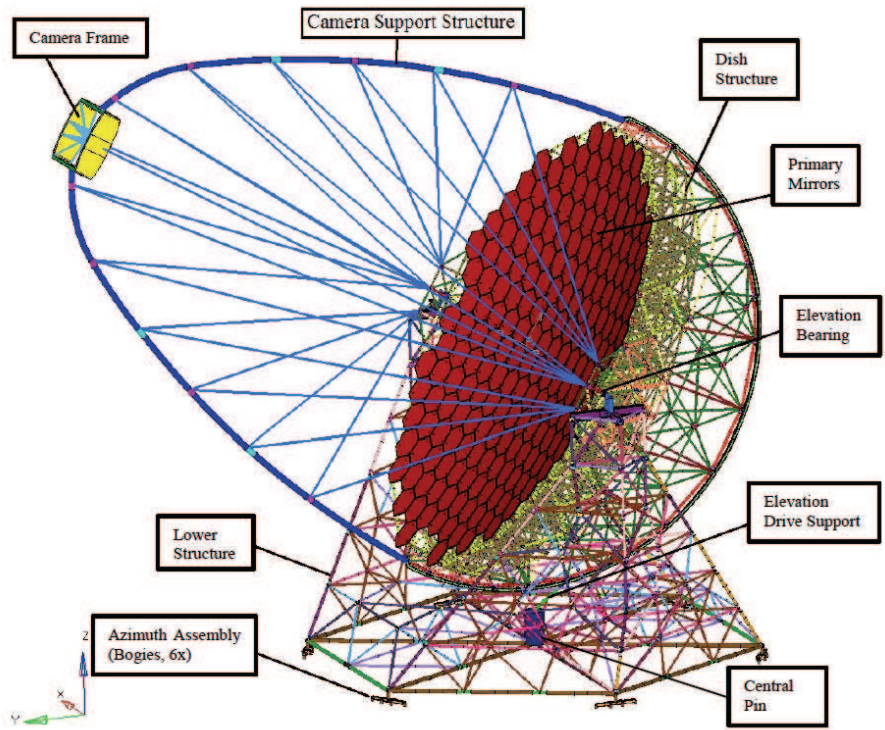}
	\caption{The LST design and its components. The camera is located 
          inside of the camera frame. 
          The final mirror layout will be slightly different to that shown here.}
	\label{fig:lst_overview}
\end{figure}

The reader is referred to the Technical Design Report (TDR) of the LST for 
a detailed description of the telescope and its component parts\cite{TDR} . 
Here we only provide a general overview of the design as illustrated in 
Figure \ref{fig:lst_overview}. The LST's main parameters are summarized in 
Table~\ref{tab:lst_params}.

\begin{table}[!htb]
	\centering
	\begin{tabular}{| l | l | c | c |}
	\hline
		\multicolumn{4}{| c |}{\textbf{LST Main Parameters}} \\
	\hline
		\multicolumn{4}{| l |}{\textbf{Optical Parameters}} \\
	\hline
				& Reflector type 			& 1-mirror, parabolic & 	\\
	\hline
	      	& Focal length 					& 28\,m 			& 	\\
	\hline
	      	& Dish diameter 				& 23\,m 			& 	\\
	\hline
	      	& f/D  				   & 1.2 			& 	\\
	\hline
	      	& Mirror area 		          & 396\,m$^{2}$	& 	w/o shadowing \\
	\hline
	      	& Mirror effective area     & 368\,m$^{2}$	& 	Including shadowing \\
	\hline
	      	& Preliminary on-axis PSF   & 0.05$^{\circ}$ & 	                \\
	\hline
	      	& Preliminary off-axis PSF   & 0.11$^{\circ}$ & at 1$^{\circ}$ off-axis 	      \\
	\hline
	      	& Preliminary tracking accuracy 	& 20\,arcsec 		& 	RMS, online precision \\
	\hline
	      	& Pointing accuracy 		& 14\,arcsec 		& 	RMS, post-calibration precision \\
	\hline
	         \multicolumn{4}{| l |}{\textbf{Camera Parameters}} \\
	\hline									
	      	& Camera dimensions (LxHxW)			& 	2.8\,m x 2.9\,m x 1.15\,m &  \\
	\hline
	      	& Weight                  			& 	$<$ 2000\,kg &  \\
	\hline
	      	& Number of pixels        			& 	1855  &  \\
	\hline
	      	& Pixel linear size       & 50 mm including light concentrator &  1.5\,inch PMT  \\
	\hline
	      	& Pixel field of view     			& 	0.1$^{\circ}$  &  \\
	\hline
	      	& Camera field of view     			& 	4.5$^{\circ}$  &  \\
	\hline
	      	& Trigger region field of view     			& 	4.5$^{\circ}$  &  \\
	\hline
	      	& Sampling speed           			& 	1 GS/s  &  \\
	\hline
	      	& Analogue buffer length   		& 	4\,$\mu$s  & for hardware stereo trigger  \\
	\hline
	      	& Readout rate            			& 	7.5\,kHz (target),   15\,kHz (goal) &  \\
	\hline
	      	& Dead time               			& 	5\% at 7.5 kHz &   \\
	\hline
	         \multicolumn{4}{| l |}{\textbf{Mechanical parameters}} \\
	\hline									
	      	& Total weight             			& 	103\,tons  & all moving parts  \\
	\hline
	      	& Repositioning speed      		& 	20\,s  & for 180$^{\circ}$ in azimuth  \\
	\hline
	      	& Elevation drive range      	     		& 	-70$^{\circ}$ to 100$^{\circ}$ &   \\
	\hline
	      	& Azimuth drive range      	     		& 	408$^{\circ}$ &   \\
	\hline
	      	& Inertia elevation      	     		& 	$\sim$6000 tons$\cdot$m$^{2}$ &   \\
	\hline
	      	& Inertia azimuth       	     		& 	$\sim$12000 tons$\cdot$m$^{2}$ &   \\
	\hline
	      	& Park position       	     	  & 	zenith angle 95$^{\circ}$ &  locked at the camera tower \\
	\hline
	      	& Height at Camera Access        & 	13\,m above ground  &  In the parking position \\
	\hline		
	\end{tabular}
	\caption{Main LST parameters} 
	\label{tab:lst_params}
\end{table}

\subsection{Telescope Structure}
The Lower Structure of the LST is made of steel
tubes; the Dish Structure is from CFRP, steel and aluminium tubes. 
The Lower Structure of the LST rests on six Bogies equally spaced in a hexagonal arrangement, 
running on
a circular Rail. Two of those Bogies are located under the Elevation Bearings, 
withstanding the higher percentage of the telescope's weight. 
The Bogies run on a circular flat Rail of 23.9\,m diameter and 500\,mm width, 
which is fixed to the Foundation through the pedestals.

The baseline design of the Camera Support Structure is based on an almost parabolic arch
geometry, reinforced along its orthogonal projection by two symmetric sets of 
stabilizing fixed headstays. Most of its elements make use of CFRP, which is well 
known to provide a very high performance to mass ratio.
On top of this arch, a square Camera Frame is mounted to hold the Camera at the 
proper location with respect to the Mirrors.

To facilitate Camera maintenance an Access Tower is placed in front of the telescope whilst 
in its parked position. A lower park position (e.g. ground-level) would require a much higher-elevation axis.
The Camera is accessed from a flat platform on top of the Access Tower.
The platform is split in two sections, which can move apart so that personnel have
straightforward access to the front- and back- sides of the Camera.

\subsection{Optical System}

The Optical System of the LST is an active optics system that includes a large parabolic reflector equipped 
with an Active Mirror Control system (find more details in \cite{icrc_optical}) and a flat focal surface.

The optical reflector is composed of 198 hexagonal mirrors each with an area of 2\,m$^{2}$. 
The total effective area including shadowing effects is 368\,m$^{2}$. The Mirrors are
manufactured using the cold slump technique 
with a sandwich structure consisting of a soda-lime glass sheet, an aluminum
honeycomb box and another glass sheet. 
The mirror box is made of stainless-steel. 
Each mirror weighs about 47\,kg and has a
drainage system to prevent any rain water from accumulating inside the honeycomb structure. 
The absolute (as measured by all reflected photons) Mirror reflectivity
between 300\,nm and 550\,nm is higher than 85\%.
The optical PSF containment diameter ($\theta_{80}$) of a single facet is less
than 1/3 of a pixel size at the centre of the Camera.

The mirrors are attached to the Dish of the LST structure using two actuators and a fixed point.
The actuators have accurate step motors
(5\,$\mu$m step size), which are controlled by the Active Mirror Control (AMC) program
to achieve the required optical performance at any moment of time. 
Each mirror facet has a small CMOS camera attached that observes a fixed-point (generated by a laser) 
on the Camera plane, and the position of the fixed-point is used as a reference to correct any 
misalignment of the mirrors. This active optics positioning is done online at a frequency of 
once every 10 seconds.

\subsection{Camera}

\begin{figure}[!htb]
  \centering
  \includegraphics[width=0.85\textwidth]{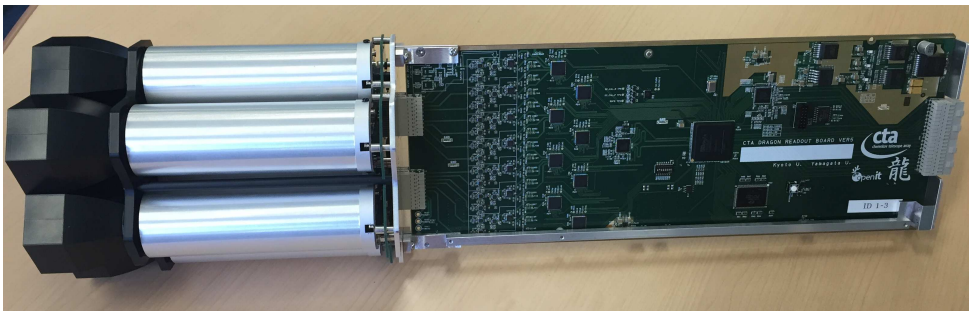}
  \caption{Photo of a \emph{PMT} Module without the aluminium shielding. }
  \label{fig:lst_single_pmt_module}
\end{figure}

The Camera of the LST shares many elements with the NectarCam \cite{icrc_nectarcam} 
proposed for use in the cameras on the Medium Size Telescopes (MSTs) of CTA.
With a weight of less than 2 tonnes the camera is comprised of 265 PMT modules 
(see Figure~\ref{fig:lst_single_pmt_module}) 
that are easy to access and maintain. Each module has 7 channels, providing the camera with a 
total of 1855 channels.
Hamamatsu photomultiplier tubes with a peak quantum efficiency of 42\%
(R11920-100) are used as photosensors converting the light to electrical
signals that can be processed by dedicated electronics. 
In addition, we continue to research the possibility of using Silicon photomultipliers 
\cite{icrc_sipm} in future versions of the LST cameras. 

For maximum light throughput each 
photosensor is equipped with an optical light concentrator, optimized for the field of
view and geometry of the photosensor \cite{icrc_light_guides}. The Camera has a total field of view of
about 4.5$^{\circ}$ and has been designed for maximum compactness and lowest weight,
cost and power consumption while keeping optimal performance at low energies.
%(see Figure~\ref{fig:camera}). 
Each pixel incorporates a photosensor as well as
the corresponding readout and trigger electronics. The readout sampling
frequency is 1~GS/s; the bandwidth of the trigger and the readout is about
300~MHz. This readout electronics is based on the DRS4 (Domino Ring Sampler
version 4) chip, which is currently used in MAGIC\cite{MAGIC}. In order to
increase the analogue buffer length, 4 DRS4 channels are cascaded. The analogue
signals are split into Low and High gains. 

The readout window is variable. Only a Region of Interest of 30~ns length at a time
position defined by the trigger signal will be read out. This window
is longer than the characteristic timescale for gamma-ray induced signals in the 
energy range of interest for the LSTs. 

The Camera trigger strategy is flexible and based on the
shower topology and the temporal evolution of the Cherenkov signal produced in
the Camera. The analogue signals from the photo-sensors are conditioned and
processed by dedicated algorithms that look for extremely short and compact
light flashes. Furthermore, the Cameras are interconnected in order to
form an on-line coincidence trigger amongst the LSTs. This enables the suppression
of accidental triggers by up to a factor of 100.

Monte Carlo simulations have shown the
Sum-Trigger to have the best performance in terms of triggering showers at the lowest energies.
Hence the Camera trigger implemented in the LST, which is aiming to reduce the energy threshold as
much as possible, adopts the Sum-Trigger configuration.

\subsection{Auxiliary systems}

The Auxiliary Systems refer to all instruments and devices on the LST excluding the Camera.
The main functions of these devices include driving the telescope, correcting the telescope pointing 
or focus and calibrating the Camera. The devices used for Structure Condition Monitoring as well as 
Lightning Protection are also considered to be Auxiliary Systems.

The fast and precise movement of the LST is achieved by using electric servomotors on
both the elevation and azimuth axis.
Four synchronized motors are used for the azimuth axis and
two for the elevation one. 
Synchronization of the motors is managed directly by the Drive controller. There are two main
strategies depending on the operational mode: fast motion or tracking. In fast motion, a regulation
method limiting the torque is used in order to ensure a good load distribution between the motors. 
While in tracking mode the speed and not the torque of the motors is limited.
The azimuth and elevation motors are located on the top of the Bogies and on the Elevation Drive Arch, 
respectively.
The Elevation Drive Arch of the Elevation Assembly is formed by a curved I-beam with a rail. This 
is for a box that carries the declination drive motors and an external brake, which in the event 
of a motor failure, prevents any Dish rotation.

The central facet position of the LST's mirror is kept vacant to accommodate several devices. 
These include: a) a Calibration Light Source box used to calibrate the gain of the camera's 
photosensors; b) an Optical Axis Reference Laser used by the AMC to define the optical axis of 
the telescope; c) an Inclinometer used to measure the pointing elevation; d) a starguider 
camera to relate the pointing of the telescope with respect to the sky field; e) a Camera 
Displacement Monitor used to monitor any sagging of the Camera; and f) Distance Meters used 
to measure any displacement or rotation of the Camera along the optical axis with respect to 
the Dish centre. This relatively sophisticated pointing setup is necessary to 
meet the strict requirements defined for the LST's pointing (14 arcsec, post-calibration) 
and lightweight structure.

\section{Current status}

Currently the telescope array layouts proposed for the CTA observatory includes eight LSTs, 
four at the CTA-North site and four more at the CTA-South site \cite{CTA}.
The working group within CTA responsible for the development of the LST has produced a plan for the 
manufacturing of the telescopes. The goal is to produce all the LSTs over a period of six years, 
starting with the installation of the first LST by the end of 2016 and ending with the 
commissioning of the last LST in 2021.

Development of the LST is different compared to the other CTA telescopes because the first LST 
constructed will also be a prototype. However, this prototype will be fully functional, and once 
commissioning finishes and verification tests successfully completed (hence demonstrating the LST 
fulfils all CTA requirements) this LST will become the first production LST of CTA (Pre-Construction). 
This decision is motivated by the relatively small number of units to be produced and the elevated 
cost of each individual telescope. If during commissioning it is found that design modifications are 
needed, the first telescope will be retrofitted to ensure its performance  is equal to that of 
the other LSTs.

The second telescope will mark the beginning of the LST Construction phase. Several elements
manufactured in workshops of participating institutes will be subcontracted to industrial partners. Orders
will be timed based on the experience obtained during the construction of the first LST, aiming for a peak
construction rate of two telescopes per year. 

\subsection{Tests}

The LST design is essentially complete\cite{TDR}, pending a very small number of outstanding 
design choices. Drafting of installation, operation and maintenance plans is ongoing. The design 
has been evaluated by an external committee of technical experts (Critical Design Review) on June 25th 2015.
During the Preparatory Phase of CTA, the LST team has prototyped, verified and validated many elements. 

{\bf LST mechanics}: It is basically an enlarged copy of the MAGIC telescopes\cite{MAGIC}. 
Most of the telescope elements used in the LST are considered proven technology. However, there are 
slight differences so a Finite Element Analysis (FEA) of the full telescope model 
is used to evaluate the design before its realization. The analysis has
been performed by means of NASTRAN FEA software and shows that the design fulfills requirements
(a possible reinforcement for the case of severe earthquakes is still under study). 

{\bf Azimuth Assembly}: Critical requirements need to be fulfilled by the Bogies. The acceleration
forces and the resulting emergency breaking forces of 0.1~rad/s$^2$ are extreme.
At the same time there are uplift forces of up to 54~tonnes due to the light-weight construction of the LST.
Part of the design has been validated using FEA. In addition the construction 
of one Bogie, one part of the Rail, 
and the tools to reproduce the real conditions is underway.

{\bf Drive}: A prototype of the Elevation Drive, based on the latest design, is currently being tested. 
In this test ring, the movement, the breaking and locking system is tested as well
as the maximal motor forces that are needed to achieve maximum acceleration required in the GRB mode.
In this setup, two 10.5~kW motors are installed and both can be operated to an overload of 4
($\sim$80 - 100~kW). This is more than the maximum electrical power peak expected for the Elevation Drive
($\sim$64~kW). Two control strategies have been implemented to satisfy both velocity requirement during
fast movement and tracking precision during observations. The implementation of these strategies
are tested with the elevation test stand, in particular the coupling between master and slave motors.

{\bf Elements of the structure}: 
The CFRP tubes (80~cm, 100~cm and 120~cm diameter) that are adopted in the LST design have already
been used and tested in MAGIC.
After considering several proposals for the Camera Support Structure an 
arch divided into a total of six curved sections has been adopted.
A prototype of a small section of the Mast has been manufactured in order to validate both the production
processes (e.g. plies lamination inside the conical shapes of the clamping part) and the efficiency of
the tapper in terms of clamping.

{\bf Mirrors}: The Mirror Facets need to be evaluated from both the optical and the 
mechanical points of view. 
Several tests are carried out in different facilities, where specific test-benches are prepared.
The mirror PSF has already been characterized and it fulfills the requirements.
The mechanical properties of the Mirror design are verified using FEA
and will be later studied with long and extensive cycles that reproduce the
foreseen strong variations of temperature, humidity and external conditions.
A segment of the dish (roughly 1/8 of the final structure) has been mounted. It has been used to test
the Mirror mounting mechanism (Interface Plates) and the Mirror installation procedure. 

{\bf Camera mechanical structure}: An FEA analysis has been performed in order to verify that it
complies with the requirements. This analysis has been performed using ANSYS software.
The FEA simulation of the PMT Modules Holder shows that the maximum deformation of less than 1 mm 
is reached in the event of a severe earthquake. 
In order to verify if the mechanical interface between the Modules Holder and PMT Modules is reliable
within the production tolerances, a prototype of the Modules Holder has been produced.

{\bf Camera cooling}: It is strongly related to the camera mechanics. Two different cooling
systems have been considered and studied: a temperature controlled air flow cooling system and a
water cooling system based on cold plates. Several simulations and tests have been performed.
Once the performance has been well established, a hybrid design, which keeps the best of both cooling
strategies, has been chosen as the common cooling design for LST and MST (NectarCAM).

{\bf Photomultiplier}: The quality control of PMTs with the Cockcroft-Walton HV supply is done 
prior to delivery by the company Hamamatsu Photonics.
For the PMTs of the first production (2000 units), the three PCBs for the HV power supply are mounted
at the base of the PMT, directly by the company.
The 2000 units of PMTs of the first production have eight dynode stages, while the PMTs of the main
production have seven dynode stages.
In order to fulfill the requirement of a pulse width shorter than 3~ns, the signals had to be attenuated 
at the input of the pre-amplifier board to operate the PMTs at higher HV values.

{\bf Readout}: The main electronic performance tests have been carried out in the lab.
The dynamic ranges of low- and high-gain channels are respectively 100 and 2000 photoelectrons (Ph.E.), 
which satisfies the design requirement specified to be >1000~Ph.E.
The measured noise level is 1.4~mV RMS, corresponding to 0.09~Ph.E. for the high gain channel.
The prototype meets the 300 MHz bandwidth design requirement that ensures high signal-to-noise for channels 
with a small number of photons.

{\bf Trigger}: The Majority and Sum-Trigger options have been tested and characterized and both
meet the specified design requirements. The final versions, ready for mass production, have been successfully tested
together with the Readout board. Working versions of the digital trigger have also been tested,
showing similar technical performance to the analog trigger designs. 

\subsection{Prototype LST in La Palma}

The first LST is essential to validate the global design. Most of its hardware elements are currently in
production and must pass the quality control specified by the LST team in 2015. 
Hundreds of photo-sensors and readout electronics
will be tested during this year. The whole Camera of the first telescope will be assembled at the end
of the year. It will be tested and characterized during the first half of 2016 before being shipped
to the site.

Since the CTA site selection has been delayed, the LST team decided to build it at a site that is easy to access and
has been proven to be suitable for Cherenkov telescopes, namely the Observatory of Roque de los
Muchachos (ORM) in La Palma, Spain. The observatory is operated by Instituto de Astrofisica de
Canarias (IAC). Most of the required infrastructure (water, roads, electricity, high-speed network and
residence for astronomers) is ready. An agreement for the installation of the first telescope between IAC
and the LST team was signed in April 2015. In May 2015 the LST team applied to the local authorities for permission 
to start civil engineering in Fall 2015. After the prototyping of the first telescope and its 
evaluation, the remaining telescopes can be installed
in the official sites of CTA. If the CTA North is selected in other place different from ORM, the first
prototyping telescope will be disassembled and moved to the official CTA site.

\section*{Acknowledgments}

We gratefully acknowledge support from the agencies and organizations 
listed under Funding Agencies at this website: http://www.cta-observatory.org/.
%This work is partially funded by the ERDF under the Spanish MINECO grant FPA2012-39502.

\end{document}